\documentclass[Afour,sagev,times,review]{sagej}

\usepackage{moreverb,url}
\usepackage{todonotes}
\usepackage{enumitem}
\usepackage{ulem}
 \usepackage{graphicx}
\usepackage[colorlinks,bookmarksopen,bookmarksnumbered,citecolor=red,urlcolor=red]{hyperref}

\newcommand\BibTeX{{\rmfamily B\kern-.05em \textsc{i\kern-.025em b}\kern-.08em
T\kern-.1667em\lower.7ex\hbox{E}\kern-.125emX}}

\def\volumeyear{2023}

\begin{document}

\runninghead{Timon et al.}
\def\journalname{Digital Health}

\title{Automatically detecting activities of daily living from in-home sensors as indicators of routine behaviour in an older population}

\author{Claire M. Timon\affilnum{1}, Pamela Hussey\affilnum{1}, Hyowon Lee\affilnum{2}, Catriona Murphy\affilnum{1}, Harsh Vardan Rai\affilnum{2} and Alan F. Smeaton\affilnum{2}}

\affiliation{
\affilnum{1}Centre for eIntegrated Care (CeIC), School of Nursing, Psychotherapy and Community Health, Dublin City University, Dublin 9, Ireland\\
\affilnum{2}Insight Centre for Data Analytics, Dublin City University, Dublin 9, Ireland}

\corrauth{Alan F. Smeaton, 
Insight Centre for Data Analytics, Dublin City University, Glasnevin, Dublin 9, Ireland.}

\email{alan.smeaton@dcu.ie}

\begin{abstract}

Objective\\
The NEX project has developed an integrated Internet of Things (IoT) system coupled with data analytics to offer unobtrusive health and wellness monitoring  supporting older adults living independently at home. Monitoring {currently} involves visualising a set of automatically detected activities of daily living (ADLs) for each participant. The detection of ADLs is achieved {} to allow the incorporation of additional participants whose  ADLs are detected without re-training the system.

Methods\\
Following an extensive User Needs and Requirements study involving 426 participants, a pilot trial and a friendly trial of the deployment, an Action Research Cycle (ARC) trial was completed. This involved 23 participants over a 10-week period each with c.20 IoT sensors in their homes. During the ARC trial, participants each took part in two data-informed briefings which presented visualisations of their own in-home activities. The briefings also gathered training data on the accuracy of  detected activities. Association rule mining was then used on the combination of data from sensors  and participant feedback to improve the automatic detection of ADLs.

Results\\
Association rule mining was used to detect a range of ADLs for each participant independently of others and  was then used to detect ADLs across participants using a single set of rules {for each ADL}. This allows additional participants to be added without the necessity of them providing training data. 

Conclusions\\
Additional participants can be added to the NEX system without the necessity to re-train the system for automatic detection of the set of their activities of daily living.

\end{abstract}


\maketitle


\section{Introduction}

Using IoT technologies, the use of ambient sensors to detect activities in the homes of older or more vulnerable people has grown in recent years \cite{10.1007/s10916-019-1365-7}.
In its basic form, the use case for this has been to record and visualise the raw data from actual sensor triggers and activations and to present aggregated views of this data spanning days, weeks or even months. This allows a clinician, a caregiver or a family member to observe whether certain sensors have been triggered or not. 
In turn it also allows an observer to use their observation of sensor activations to deduce whether or not higher level activities to do with eating, cleaning or social interaction with others, have taken place. For example if IoT sensors on the kettle and on the doors to the cupboards where cups, tea and sugar are stored are all activated within a short time frame during the morning, then the observer could infer that a mid-morning tea or coffee was made.

\begin{figure*}[htb]
\centering
\includegraphics[width=\textwidth]{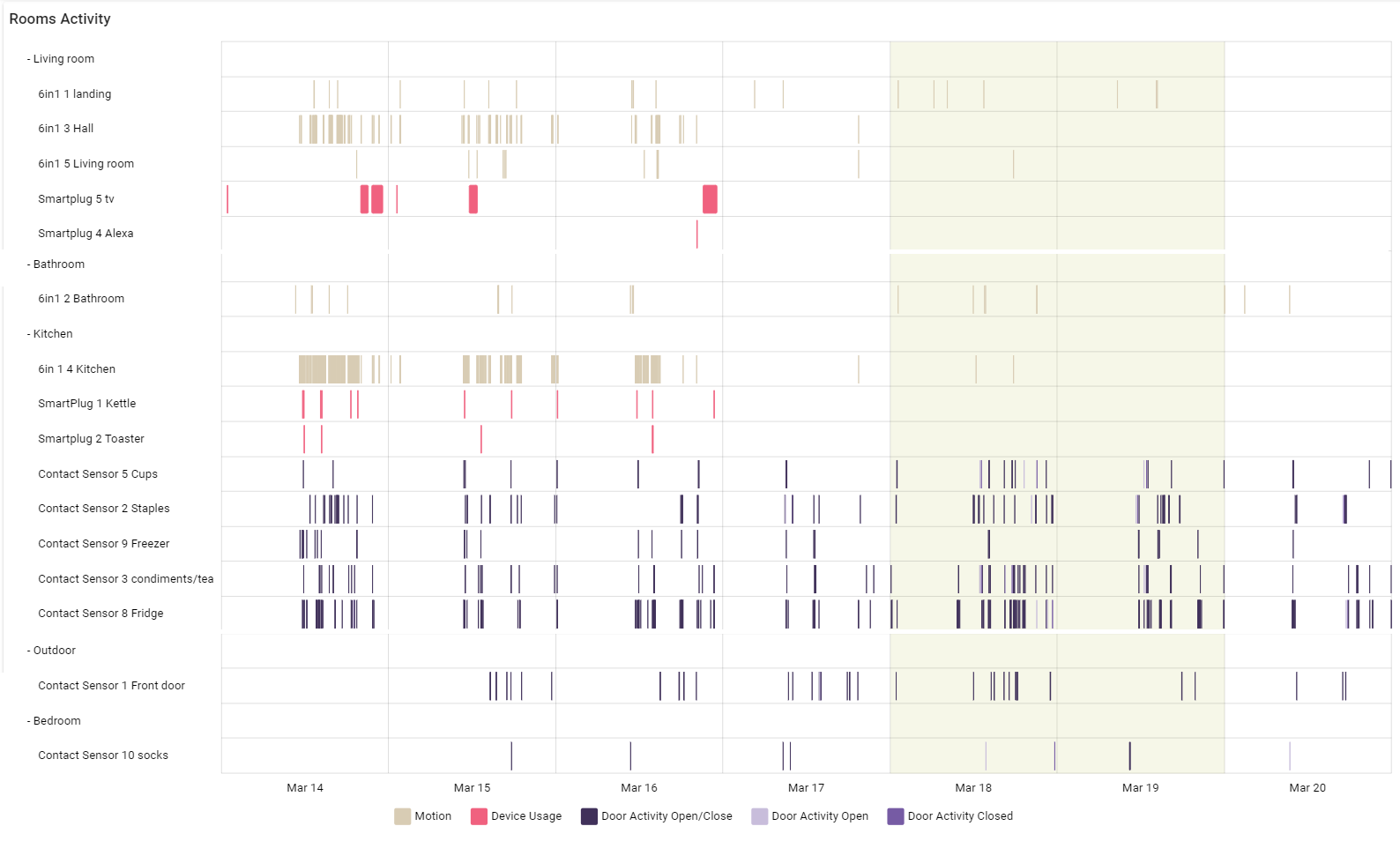}
\caption{Visualisation of raw sensor data from a participant's home showing activations of 16 different sensors  categorised into motion, environmental, electrical device use and contact sensors (y-axis). The data visualisation spans 7 days (x-axis) with the weekend (March 18th and 19th) highlighted.  \label{fig:raws}}
\end{figure*}

Visualising raw data from sensors can allow patterns of in-home behaviour to be observed but this is far more challenging because typically there are a large number of sensor activations that are not connected with the higher level activities {which} we may wish to observe as well as the general visual ``noise'' from visualising so much data. For example, just because a sensor on the entrance door to a home has been activated does not mean the occupant has left or arrived, the activation could have been caused by a caller to the home, or by a delivery. It is only by looking at  combinations of sensor activations {in occasions of ADL activities} that the actual behaviour can be accurately determined. So if presence sensors in more than one part of the home are simultaneously  activated after the entrance door sensor has been activated that implies there is a caller to the home.

While the approaches to gathering and visualising raw sensor activations are useful, their limitation is that they place the burden on the clinician or observer to interpret  raw sensor activations into higher level activities which correspond to the things that people do in their everyday life by grouping combinations of sensor activations.
This can be seen  in Figure~\ref{fig:raws} from our deployed system showing one week of raw sensor data from a participant's home. A total of 16 sensors are deployed including motion sensors, 6-in-1 environmental sensors, smartplugs and contact sensors on doors and presses. While scanning this visualisation can reveal daily daytime and evening patterns of activities particularly in the kitchen and other rooms, it is difficult to get an overall view and especially to extend an overall view of activities into multiple weeks

Activities of Daily Living (ADLs) are a set of known, pre-defined and agreed  daily physical or movement activities which most people will carry out and which correspond to the skills required to manage our basic physical needs \cite{ADLs}. Proposals for what make up a definitive set of ADLs have been around for many years \cite{lawton1969assessment} and some have been revised since those first proposals and specialised for  areas 
including activities for people with dementia and activities for stroke patients \cite{nouri1987extended, galasko1997inventory, hindmarch1998bayer}.  
Even with such subject specialisms, the set of ADLs commonly used today are fairly stable \cite{ADLs}. 

ADLs are typically used to provide a summative assessment of whether a person is able to reach a certain level of movement and to competently complete basic tasks so self-manage their lives and typically this would be used in assessments of older citizens \cite{doi:10.1177/20552076221084468}.
ADLs are essential and routine tasks that most healthy individuals can perform without assistance \cite{lawton1969assessment}. The inability to accomplish essential activities of daily living may lead to unsafe conditions, poor quality of life and may be indicative of a physical or cognitive disability in older adults. Eligibility for home care is frequently associated with deficits in ADL ability \cite{kemper2008meeting,rothgang2003dependency}. Assessment of ADLs through self-reported, observed or objective data provides evidence to individuals and caregivers on existing baselines and potential deficits in self-care ability and supports potential interventions which may be required for continued independence. 

{The state of the art in the field of recognition of activities of daily living is already well developed as shown by  systematic reviews published within the last decade including \cite{rashidi2012survey, reeder2013framing, queiros2015usability, blackman2016ambient, cicirelli2021ambient}. These works describe a field which has received much attention because it is an important topic and it has a very practical and useful nature.}

In this paper we present a technique to automatically detect a subset of common ADLs from raw sensor data in the homes of older citizens living alone and to ``tag'' their routine behaviour. The sensors used in our study of ADL generation are not wearable sensors but are in-situ sensors in the home though participants did use a smartwatch which was not used in this study. The set of ADLs are chosen as indicators of routine everyday behaviour. The ability to infer and visualise higher level activities as well as {viewing} the raw sensor data means that caregivers and family, as well as participants themselves, can assess behaviour and behaviour changes over time in a more natural and intuitive way.  

The technique we use for inferring ADLs uses association rule mining and relies on an initial set of manual annotations from participants but once this is in place we can incorporate additional participants without the necessity for further manual annotation.
{While our approach to ADL detection is data-driven, other approaches to ADL detection have been taken including a knowledge-driven approach in \cite{chen2011knowledge}  which uses domain knowledge, structured ontologies and semantic reasoning to disambiguate potential conflicts. The focus of the work in \cite{chen2011knowledge} is on real-time detection of ADLs as they happen, in an incremental way hence the use of semantic reasoning and ontologies to disambiguate. In the work in this paper the detection of ADLs happens retrospectively, at the end of each day because our use case does not require real-time detection.}

{The work which is possibly closest to what we report here is a series of works by researchers from INRIA in France \cite{caroux2014verification,caroux2018towards}. Their work involved several older healthy participants, living normally in their homes and targeting a range of daily activities to detect while using sensor data to assist in the detection.  That work 
culminated in a method to detect 6 generic activity types including meal preparation, leaving the home, and dressing/waking up which overlap with the ADLs we use in this paper, and was tested on 5 adults over a short period of 5 days \cite{belloum2021tooled}.
In that work the target was activity verification where the participants' declarations of their own daily activities were refined with sensor logs and visualised for them for confirmation. The work we report in this paper targets detection of similar activities of daily living but we take a more data-driven approach, are less reliant on participants' self-verification of their activities and our experiments are larger with more participants and over a longer period of data logging.}


\section{Methods} 

The overall aim of the Action Research Cycle Trial (ARC) trial was to investigate the technical performance and participant evaluation of a refined version of the NEX system. Ethical approval to conduct the ARC trial was obtained from the Dublin City University Research Ethics Committee (DCUREC202221) on 25/1/2022. The NEX ARC Trial was advertised through various networks including the Age Friendly University, Dublin City University and NEX study social media platforms. Eligibility criteria to participate in the trial included: demonstrated capacity to provide written consent as determined by a cognitive assessment \cite{borson2000mini}, willingness  to provide written informed consent to participate, aged 60 years or over, with or without one or more stable chronic condition/s,  fully vaccinated against COVID-19 and had an active Wi-Fi connection at home. Older adult participants were enrolled to the study for a 10-week period if they met the eligibility criteria.

Between January 2022 and July 2022, twenty-six healthy older adults (aged 60 years and over) who were living independently at home in the community participated in the trial. The gender profile was predominantly female (81\% n=21) with a total population mean age of 73.2 years. All participants resided in Dublin, Ireland (100\% n=26) and the majority lived in urban locations (96\% n=25). This was a well-educated sample as 65\% (n=17) received third level education. The majority of participants within this sample present as independent and high functioning as only 8\% (n=2) reported difficulties in completing activities of daily living (ADLS) such as dressing etc. and only 4\% (n=1) reported difficulties in completing more complex tasks defined as instrumental activities of daily living (IADLS) such as shopping for groceries etc. Three participants dropped out and one participant was no longer able to stay involved with the trial as her Wi-Fi connection was deemed too weak to support the NEX system on inspection by the technical engineer during a site visit, resulting in a final sample of n=22.

The research team devised a study design which greatly minimised face-to-face contact with participants in an effort to minimise the risk of COVID-19 spread. This meant that the majority of study visits were completed over Zoom. After enrolment to the trial, participants met with a researcher on Zoom to complete a demographics questionnaire, a questionnaire about technology use, and a compilation of health and well-being assessments. Additionally during these research calls, the researcher completed a home configuration assessment in collaboration with participants. The purpose of this home configuration was to inform the research team about the participant’s home layout and  their routine so that decisions about the appropriate placement of IoT sensors and smart plugs could be made. The assessment consisted of a number of questions e.g. the type of home where the participant lived, number of rooms, number of external doors, doors used most often, the layout of the participants' kitchen, which cabinets were used to store food, what appliances were used most frequently, etc.

During a second visit, a researcher and technical engineer visited the participant in their home environment to facilitate the installation of the NEX system technology. The researcher, technician and participant complied with a very strict COVID-19 study protocol which was developed by the research team and consisted of antigen testing prior to,  and mask wearing during, home visits. The researcher and technician used home configuration assessment with the participant in Visit 1 to determine the most appropriate placement of preconfigured technology. The NEX system design consisted of a range of IoT technologies, including a smartwatch (for measurement of sleep and step count), voice activated assistant (entertainment and reminder functionality), contact sensors (detecting activity around the home and opening and closing of doors and cupboards), smart plugs (measuring energy use of appliances), motion sensors (detecting movement, temperature, humidity, and light in the home), hub (a central connection point for sensor devices), tablet (display NEX system data to participants), and a cloud hosted secure device management platform.

The technologies were deployed in combination to facilitate the detection of some of the key ADLs from  participants’ {in-home} sensor  and smart plug use data over the trial period. Face-to-face training on the technology was provided to  participants at the time of installation, and a training manual and a series of training videos were also provided.  Throughout the remainder of the ARC trial the researchers met with {19 of the 23} participants individually on two subsequent occasions over Zoom {and met with the other 4 once,} to present them with a snapshot of raw data that was collected via the NEX technologies in the previous 24 hours. In preparation for these, sensor data for each participant was pre-processed to generate candidate occurrences of ADLs. These were presented to participants for validation and the briefings also included gathering recollections of in-home activities in the day or days immediately preceding the briefings e.g. confirming What time did they have breakfast at? etc. These provided  training data for subsequent ADL detection.  

At the end of the ARC trial the technical engineer visited the participant in their home and removed all of the technology. During the final research visit, the researcher interviewed participants about their experience of the trial and the NEX technology and completed an assessment of the system acceptability and usability (adapted version of the Technology Acceptance Model \cite{davis1989user} and System Usability Scale \cite{brooke1996sus}. The researcher also repeated the health and wellbeing measures administered at the start of the trial to investigate whether having NEX installed in participants’ homes for the duration of the trial affected their wellbeing and other aspects of life.

While there are {many} individual ADLs we could focus on, we balanced the value of different ADLs given the characteristics and demographics of the ARC trial participants against the feasibility of detecting ADLs  given the  sensors which were  deployed in their homes. After much consideration  {and taking the requirements of the clinical partners into consideration}  we focused on  4 ADLs and {grouped} each with a set of  in-home sensors which could  be used to detect them automatically. These ADLs are presented in  Table~\ref{tab:sensors}. {Increasing the number of ADLs would not affect the validity of our approach since each additional ADL would be grouped with a set of sensors needed to detect it and each additional ADL would have its own set of rules for detection. Table~\ref{tab:sensors} shows that the sets of in-home sensors used for each ADL in this work do not overlap but even if they did, that would not affect the performance of ADL  detection.}

\begin{table*}[htb]
\centering
\begin{tabular}{lp{10.5cm}} 
 \toprule
 Activity of Daily Living &	 In-home sensors likely to detect the ADL \\
 \midrule
Eating or drinking &
Contact sensors on cupboard or drawer doors for crockery, cutlery, delph, staples, pots, the fridge door and
plug sensors on the kettle, microwave and toaster; \\
 \midrule
Dressing&	Contact sensor on wardrobe door(s) and on  drawer(s);\\
\midrule
Bathing	&6-in-1 sensor, including humidity, in the bathroom; \\
 \midrule
Leaving the house&	Contact  sensor on front door, patio door or back door; \\
  \bottomrule
\end{tabular}
\caption{Mapping between ADLs and the in-home sensors likely to be used to detect them.}
\label{tab:sensors}
\end{table*}

To turn the training data for ADL occurrence into automatic detection of ADLs we examined different machine learning techniques that could be used to build  classifiers to recognise ADLs.  
Within the field of machine learning, deep learning approaches are regarded as best in terms of accuracy but their downside is that they need  much training data in order to be reliable \cite{zhang2021understanding}. In addition, once the models have been created they cannot offer any explanation for  recommendations or outputs that they generate \cite{gilpin2018explaining}.  Our application  has  limited amounts of training data because there are only so many times we can ask participants to indicate when they had eating, sleeping, bathing or other ADL activities before user fatigue sets in and the quality of the annotations deteriorates.  Our participants and our clinical partners are also wary of black box machine learning precisely because they have no explanation capabilities.

Association rule mining (ARM) is a machine learning technique which automatically develops conditional rules based on input data such as sensor data readings and annotated training data \cite{solanki2015survey}. It is a technique which has been around for many years and used successfully in a wide range of applications \cite{10.5555/3000292.3000305}.

As the name implies, association rules are a series of  if/then statements that aid the discovery of relationships between seemingly unrelated data collections. ARM seeks to identify recurring patterns, correlations, or relationships in  datasets. A rule generated by the ARM process has two parts, and antecedent and a consequent.  An item found in a data collection is called an antecedent, and an item found in combination with an antecedent is called a consequent. For instance consider the following:

``A participant is 90\% more likely to watch television when he/she is having breakfast."

In this case, breakfast is the antecedent and watching TV is the consequent in the association rule above. 

The process of developing sets of association rules involves carefully reviewing  data and searching for recurring if/then patterns. The most significant associations are then determined according to the following two parameters:
\begin{itemize}
    \item $Support$ which describes how frequently the data collection contains instances of the if/then relationship;
    \item $Confidence$ which is the number of times these associations have been verified to be accurate iin the data collection.
\end{itemize}

\noindent 
When processing large datasets using association rule mining, for every conceivable item combination of data items, the Apriori algorithm \cite{al2014improved} is attractive to use as it scans the data collection only once as it  derives a set of association rules.  In an earlier phase of our work we  validated that the Apriori algorithm can  be used successfully to detect ADLs using the data from 7 participants in a friendly trial where we detected kitchen events only \cite{orla}.  The results from the earlier trial indicated that for a given participant we could mine rules for the occurrence of kitchen-based activities  if we have training data for occurrences of those activities from  data-informed briefings.  

In practice the requirement for having to have training data for ADL occurrence is not scalable to larger sets of participants so our aim in generating ADLs in this work is to use the  annotations from briefings with  ARC participants and apply them unseen to  new participants.  This consideration also influenced our choice of using association rule mining for ADL detection in the ARC trial.

Processing with the Apriori algorithm for association rule mining required setting  minimum values for the $support$ and $confidence$ variables. This should indicate that we are only interested in discovering rules for things that have a minimum value for co-occurrence with other items and have a specific default existence. In this work these values have been set as $min\_support=0.15$ and $min\_confidence=0.5$.

Detecting relatively  short-duration activities of daily living requiring a small number of   activations of a dependent set of sensors but not in a particular sequence, presented challenges in the temporal domain. For example a participant may take a longer or shorter time to complete any of the ADLs and may activate sensors in a different order each time, for example putting the kettle on first and then preparing the crockery in the morning, and then doing these in the reverse order the  in the afternoon.   
To address this we used sliding windows  to aggregate  sensor activations over a set time period of various durations, effectively grouping the sensor data into an order-independent  set and thus smoothing out  variations in the ordering.

It was crucial to choose a window size that was both small enough to detect individual activities and large enough to reduce noise associated with smaller window sizes.
Analysis of the data-informed briefings with participants provided insights to establish a baseline for the size of the sliding windows for various ADLs and combined with experiments reported later, the window sizes chosen were as shown below:
\begin{itemize}
    \item For ‘Dressing’ and  ‘Leaving House’ the window sizes were 30 minutes;
    \item For ‘Eating/Drinking’ and ‘Bathing’, the window sizes were 60 minutes.
\end{itemize}
{The shift or stride for ADL detection was set to 5 minutes. That means that the association rules test for the presence of an ADL in a given time window (30 or 60 minutes) and if not present then the window would shift forward by 5 minutes and would re-test.}

{Our choice of 30 or 60 minutes for window sizes is in line with the work in \cite{belloum2021tooled} where those authors use window sizes varying from 30 to 120 minutes for the same ADLs as we detect here, though their windows begin and end at fixed times and theirs do not slide and overlap as ours do.  Our method  accommodates ADLs taking place close to each other in time because each ADL detection runs independently of others. Thus our approach will detect a participant dressing directly after taking a bath, for example. This can be seen later  in Figure~\ref{fig:ADLs}  where ADL co-occurrences are shown to overlap for a participant.}

With the window sizes for ADLs determined and using the training data from participant briefings, association rules for ADL detection from sensor data were generated, initially for each {ADL for each} participant.  To illustrate, {the conditions for some ADL detection rules for  participant 11 are shown in Figure~\ref{fig:ADL-Rule}. These show that for an Eating/Drinking event the use of any of the kitchen appliances or opening of the doors to the food staples, combined with presence detection, is the trigger. For the Bathing ADL, detecting presence and an increase in humidity within the time window is the trigger while for the Dressing ADL,  opening the wardrobe and the underwear drawer is the trigger, for this participant.
If these sensor activation conditions are satisfied within a 60-minute or within a 30-minute time window depending on the ADL, the activity will be labelled as that ADL.}

\begin{figure*}[htb]
\begin{verbatim}
            # Eating/Drinking Event
                ((Dataframe['Kitchen_sensor'] == 1) AND
                 (Dataframe['Toaster']== 1) OR
                 (Dataframe['Fridge']== 1) OR
                 (Dataframe['Kettle']== 1)  OR
                 (Dataframe['Kitchen']== 1) OR
                 (Dataframe['Microwave']== 1)  OR
                 (Dataframe['Staple']== 1) OR
                 (Dataframe['Staple2']== 1)
                )   

            # Bathing
                ((Dataframe['BathroomH']== 1) AND 
                 (Dataframe['BathroomM']== 1)
                )
 
            # Dressing
                ((Dataframe['Underwear']== 1) OR
                 (Dataframe['Wardrobe']== 1)
                )  
\end{verbatim}

\caption{Example of {conditions for some} of the association rules generated to determine some ADLs {for participant 11.}\label{fig:ADL-Rule}}
\end{figure*}

For creating groundtruth training data, the clinical partners  met with each of the participants on at least 1 occasion after the sensors had been installed in their homes for a briefing or a data-informed recall on how their deployment was going. During these  meetings held over Zoom because of the pandemic, the clinicians gathered data on occurrences of the 5 ADLs that had happened in the previous days, noting the ADL and the timestamp and this recall was prompted by the clinician sharing a visualising of the raw sensor data with the participant on the SafeTRX platform. So seeing sensors for, say, the kitchen being activated in mid-afternoon would prompt the participant to remember that s/he had made tea and had a biscuit during that afternoon which would be recorded as an eating or drinking ADL. 

\begin{figure*}[htb]
\centering
\includegraphics[width=\textwidth]{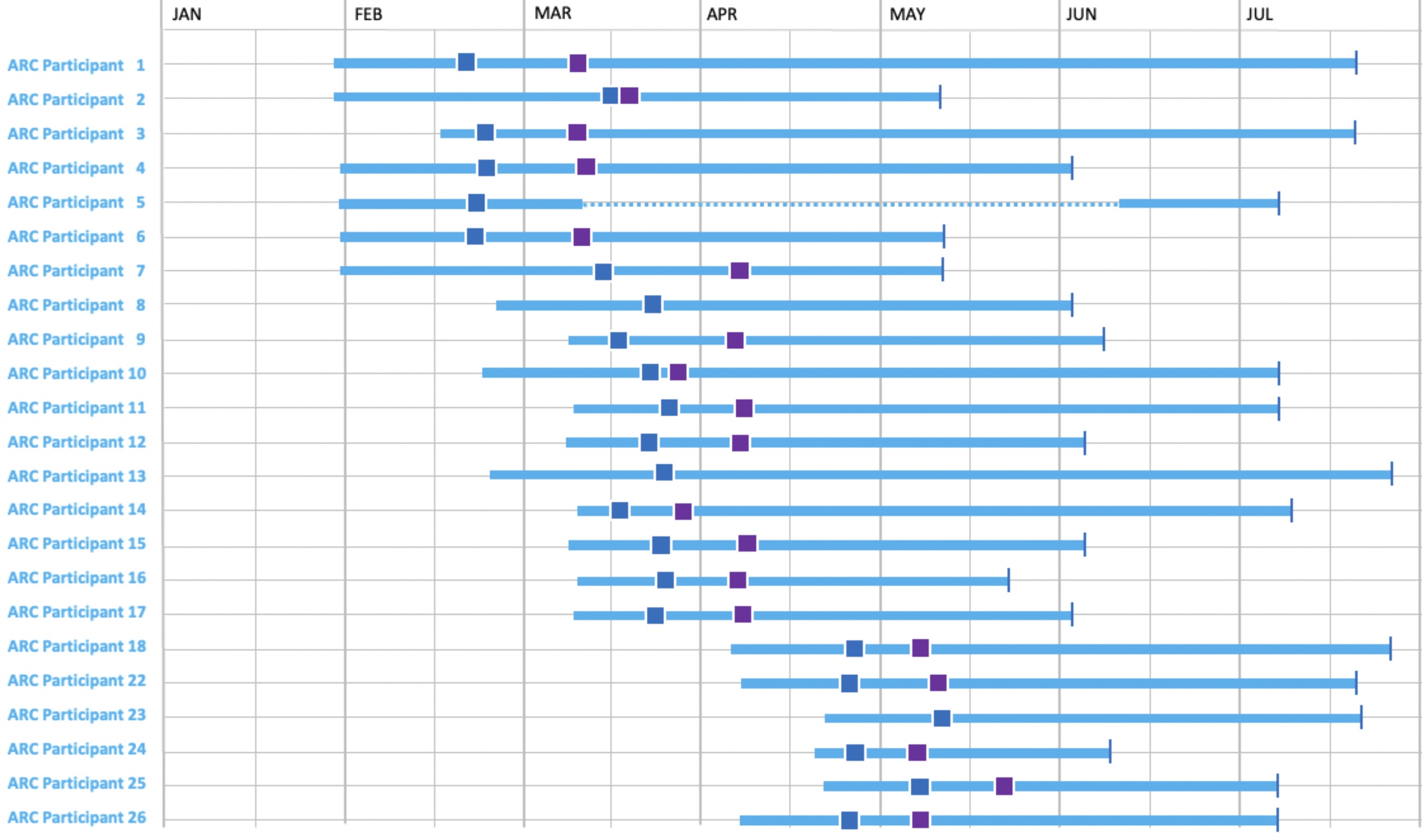}
\caption{Logging period for ARC trial participants - the blue square 
shows the first data-informed briefing, the purple square 
shows the second data-informed briefing.}
\label{fig:logging}
\end{figure*}

The timings of the clinical partner’s data-informed briefings, and their place in the overall data logging for the ARC participants is shown in  Figure~\ref{fig:logging}  which shows sensor data logging for 159 days from 23 participants.  Here we see that  23 participants, all except participants 5, 8, 13, and 23 had two briefings and that the briefings were rarely on consecutive days and most were at least one, and closer to two, weeks apart.  

\section{Results}

We developed a number of versions of using  association rule mining to build sets of rules to detect  ADLs. This was so we could (1) incrementally determine the best time window sizes to use for different ADLs and (2) include more of  participants' validations of candidate ADLs and their suggestions of additional ones from their data-informed briefings.  Different versions of the rule mining generated  different sets of ADLs for the same participants. It was necessary for the rule generation and the subsequent ADL detection to take place immediately prior to participants having one of their briefings  so that some of the candidate ADLs could be presented to them during their interviews.  

We started our use of association rule mining using participants' feedback from their first briefing with no candidate ADLs offered to them as we had no training data, and treating each participant independently of others. For their second briefing we offered  candidate ADLs generated using training data  from their first briefing and these were validated and further training data was gathered during the second briefing. {As mentioned earlier, we experimented with varying the sizes of time windows for different ADLs choosing 30 and 60 minutes depending on the ADL and
generating ADLs for each participant based on their own set of rules, independent of others.} Finally we used association rule mining to generate a single set of rules for ADL detection which we applied across all participants.
Note that not all participants were used for ADL detection at the all stages of the investigation  depending on the timing of their  briefings and the availability of their own sensor data as shown in Figure~\ref{fig:logging}.

As mentioned above, different deployments of association rule mining generated different  ADLs raising the question of whether a new set of activities is better than a previous one. Evaluating the effectiveness of a set of rules can only be done by validating the ADLs it generates against manually annotated training data, to which we have no further access, and we cannot go back to participants to get this. {This is a consequence of our focus to have little  annotation data from participants from their data-informed briefings which we could use as ground truth for training ADL detection and/or for evaluation of different ADL detection rule sets.} Thus our evaluation is done in terms of how the distribution of ADLs generated by a version, appears overall.

\begin{figure*}[htb]
    \centering
    \includegraphics[width=0.7\textwidth]{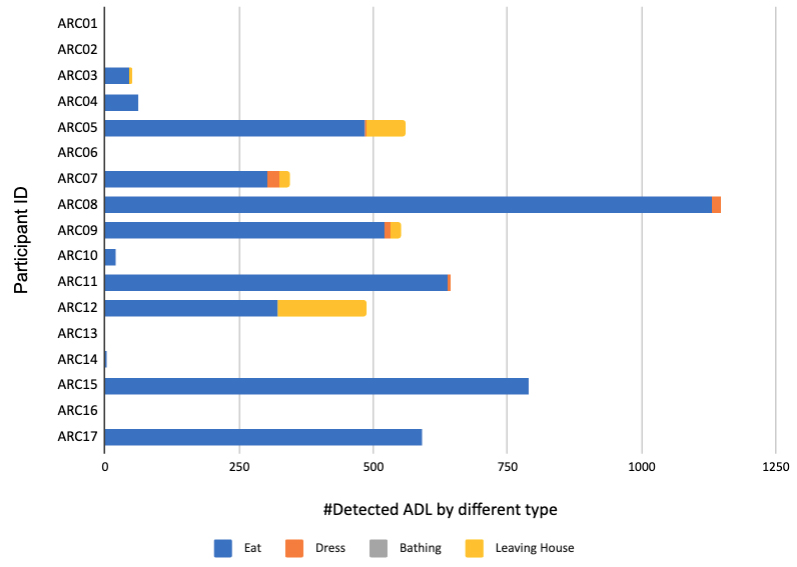}
    \caption{ADLs detected from an early version of Association Rule Mining, {showing the raw counting of all detected ADLs (x-axis) stacked by different ADL types for each of the participants (y-axis)}}
    \label{fig:V2-ADLs-raw}
\end{figure*}

\begin{figure*}[htb]
    \centering
    \includegraphics[width=0.7\textwidth]{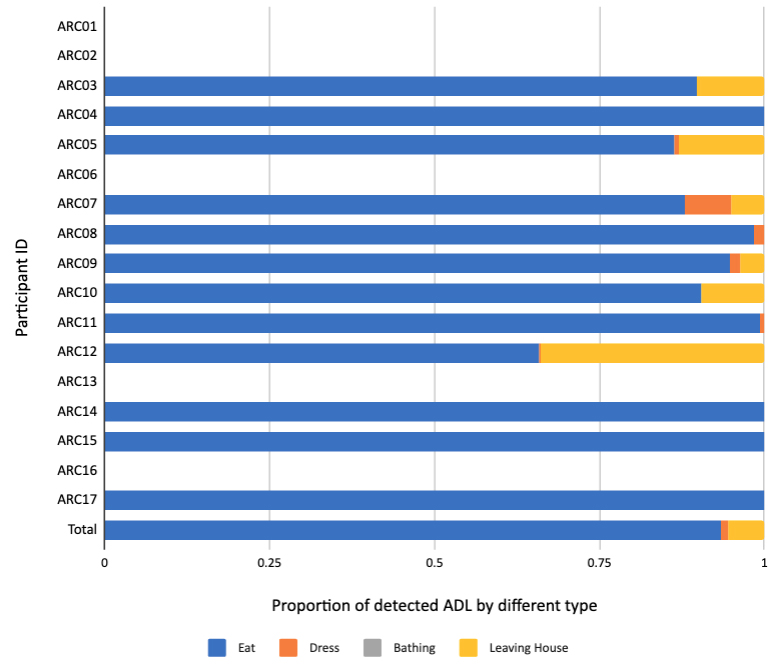}
    \caption{Proportions {(between 0 and 1)} of ADLs detected from an early version of Association Rule Mining {(proportional view of Figure~\ref{fig:V2-ADLs-raw})}}
    \label{fig:V2-ADLs-prop}
\end{figure*}

An early version of our rule mining is where training data has been generated from 2 data-informed briefings and where each participant’s ADL generation is completed independent of others but with no adjustment of window sizes for  resolution of clashing ADLs for the same participant. Figure~\ref{fig:V2-ADLs-raw}  shows the raw number of ADLs of each type  where eating ADLs dominate, and  ADLs have not been generated for all participants at that point because not all ARC installations had been completed. The different numbers of (absolute) ADLs for different participants reflects the fact that participant data logging had been running for different durations for different participants. Figure~\ref{fig:V2-ADLs-prop} shows the proportion of ADLs types for participants for this ARM version and that is a more useful indicator. From this we can see that the leaving house and dressing ADLs was not detected for some participants and the bathing ADL was not detected for any because the humidity sensor in the 6-in-1 bathroom sensor was sampling once every 10 minutes which was insufficient but subsequently corrected.

After several iterations of  association rule mining development, our final implementation  generates ADL rules from the training data from all participants, uses the optimal window settings for different ADLs and resolves clashes and overlaps between ADLs.

\begin{figure*}[htb]
    \centering
\includegraphics[width=0.7\textwidth]{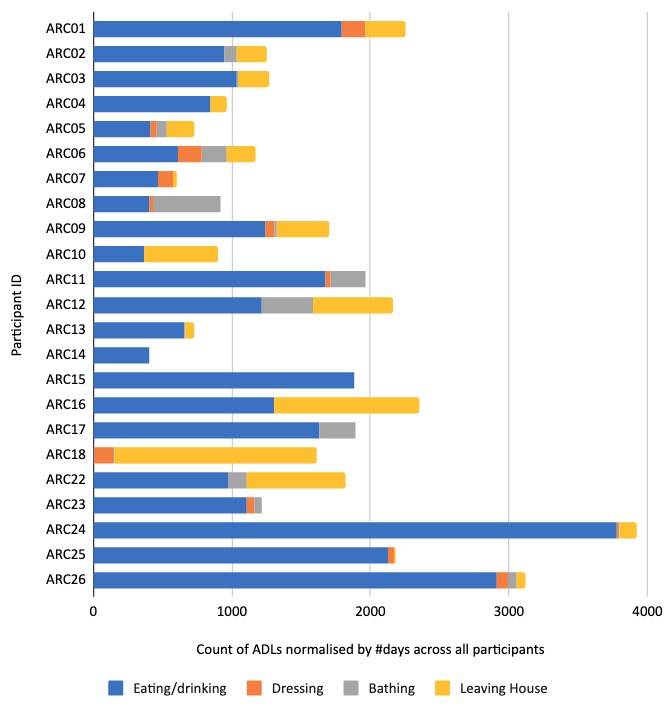}
    \caption{ADLs detected from final version of Association Rule Mining, {normalised by the number of days of data logging per participant which is taken from Figure~\ref{fig:logging}, showing the  counts of  detected ADLs (x-axis) stacked by  ADL type for each participant (y-axis)}}
    \label{fig:V5-ADLs-raw}
    \label{fig:V5-ADLs-Day-Normalised}
\end{figure*}

\begin{figure*}[htb]
    \centering
    \includegraphics[width=0.7\textwidth]{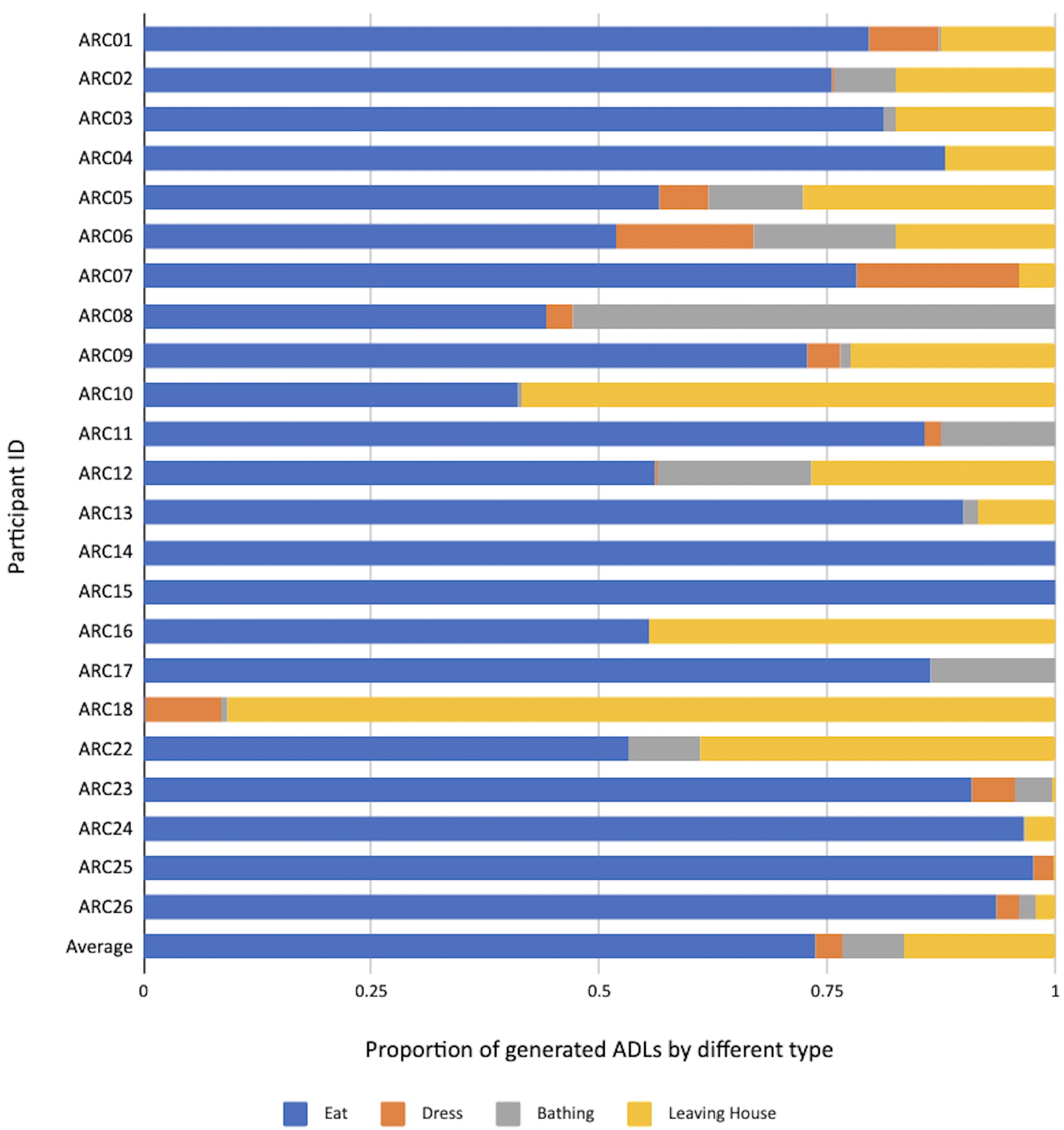}
    \caption{Proportions {(between 0 and 1)} of ADLs detected from final version 5 of Association Rule Mining {(proportional view of Figure~\ref{fig:V5-ADLs-raw})}}
    \label{fig:V5-ADLs-prop}
\end{figure*}

{Figures~\ref{fig:V5-ADLs-Day-Normalised} and \ref{fig:V5-ADLs-prop} show ADLs generated for all ARC participants.  In Figure~\ref{fig:V5-ADLs-Day-Normalised} the numbers of ADLs per participant are normalised by the total number of days by ARC01 (171 days) taken as the longest duration of all participants for the data capture in this study.}
{The normalised view helps us draw comparisons across participants for their relative {\em amounts} of ADLs, given their numbers of ADLs are for the same logging period.  In Figure~\ref{fig:V5-ADLs-prop} we show the relative {\em proportions} of ADLs per participant. The figures show} some outliers and errors like no ``eating or drinking" ADL and a disproportionally high number of ``leaving house" ADLs for ARC18 and some participants with no dressing ADL. 

{When results from all participants were completed we analysed each participant's ADLs individually with the clinician who carried out their data-informed briefing.}
For each of the outliers and errors in ADL detection were were able to determine an explanation such as having no training data to work {with for a given ADL  from the online participant briefings such as a participant not leaving their home recently} or having no appropriate deployed sensors for an ADL for a given participant.  

\begin{figure*}[htb]
\centering
\includegraphics[width=\textwidth]{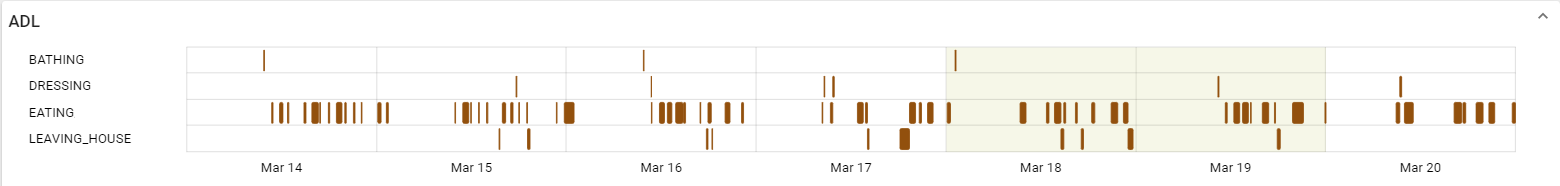}
\caption{ADLs automatically detected from raw sensor data shown in Figure~\ref{fig:raws} using association rules mined from  training data from all participants. \label{fig:ADLs}}
\end{figure*}

{The numbers of detected ADLs across participants in Figure~\ref{fig:V5-ADLs-Day-Normalised} does show a lot of variety. ARC24 shows largest number  because of the large number of eating events, similar to ARC26 and is explained as follows.}
Figure~\ref{fig:ADLs} shows the ADLs generated for   participant ARC24 over the same time period as the raw sensor data shown earlier in Figure~\ref{fig:raws}. This shows a regular bathing and dressing activity and a leaving of the house on 5 of the 7 days. March 20 shows the participant not leaving the house though the front door was opened and March 15, 16, 17 and 18 show a lot of front door activity not identified as leaving the house so the participant must have had callers or deliveries.  The eating activity is well represented throughout each day {because as shown in Figure~\ref{fig:raws} this participant does seem to spend large parts of the day in and out of the kitchen, opening and closing the fridge, food presses and drawers. Some of these recognised as the eating/drinking ADL may actually be food preparation  or returning from grocery shopping rather than food or drink consumption.}

{Other observations from Figures~\ref{fig:V5-ADLs-Day-Normalised} and \ref{fig:V5-ADLs-prop} show a high number of leaving the house ADLs for some participants, especially ARC18. This can be traced back to the fact that ARC18 had more callers to the front door than others.}

{As part of the analysis of each participant's ADLs  with the clinician who carried out their data-informed briefing as mentioned above, we analysed which of the in-home sensors appeared most often in the rules and which were used most in the triggering of those rules. From this analysis we identified} 11 core in-home sensors which should be included for any new participants for whom automatic detection of these ADLs is desired. This set of 11 is driven by their common use across all our ARC participants, and their use in the rule mining for ADL recognition and assumes that the same ADLs are the target for detection.  The 11 core sensors are listed in Table~\ref{tab:core-sensors}.

\begin{table*}[htb]
    \centering
    \begin{tabular}{ll}
    \toprule
    Sensor type & Location \\
    \midrule
    Plug-in electrical & Microwave, kettle, toaster \\
    Contact sensor & Front door, fridge door, doors to delph, cutlery and staples, wardrobe doors \\
    6-in-1 multi-sensor~~~~~~~~ & Bathroom and kitchen \\
    \bottomrule 
    \end{tabular}
    \caption{Core in-home sensors for detection of ADLs}
    \label{tab:core-sensors}
\end{table*}


\section{Conclusions}

This paper describes the data  processing carried out on in-home sensor data gathered from 23 participants over periods varying from 6 weeks to 6 months.  The  sensor data was processed into a set of activities of daily living (ADLs) which were chosen as typical indicators of regular, routine behaviour by the participants.  A characteristic of our use case of turning sensor data into ADLs is that there is a  limited amount of training data available. Our training data was gathered directly from participants during two online data-informed briefings and corresponds to participants indicating, or validating, an instance of an ADL occurrence as being true and valid.   We then used this as input to association rule mining to determine a set of rules for ADL detection.

Our initial sets of ADLs were based on a different set of association rules for each participant and then we fused the training data to generate a set of rules for detecting ADLs across all participants. This means that we can now add additional participants without requiring additional training data by re-using the training data from the pool of 23 ARC trial participants.  In this way our ADL detection is scalable and can be made available to others.

{One of the unresolved questions about the work in this paper is the end-goal and what to do with detected ADLs. In a clinical setting even the visualisation of ADLs over time has limited capacity to support observations of subtle behaviour changes and degradations. In our future work we will apply the automatic detection of periodicity intensity, namely how strongly or weakly the activities of a participant fits into the regular 24-hour circadian rhythm or the weekly cycle of behaviours, to detected ADLs. It is known that strong rhythmicity in our lives is an indicator of wellness and that degradations in our regular behaviour can be detected automatically as weakening of the strengths of our circadian and other regular rhythms.  We have already done this work using raw sensor data as input to periodicity detection \cite{smeaton2023} but believe that using higher level ADLs will give even better detection of behaviour changes.}

{There are also limitations including the limited number of ADLs (n=4) especially since there is work elsewhere reporting detection and use of larger numbers of ADLs. However our aim was to demonstrate that our technique for ADL detection with limited training data and a limited number of sensors per participant works and can be applied to new participants without the need for additional training data and with acceptable accuracy. This has been demonstrated for 4 ADLs whose detections work independently and in future work we will examine the accuracy of ADL detection when using only the 11 core sensors we identified for future deployments.  We also acknowledge that the approach could be improved with further inputs from the caregivers or directly from the participants in their homes as a form of human-in-the-loop active (machine) learning \cite{monarch2021human} where the rules would evolve and improve as more annotations were provided.}

{In summary, t}he work reported here has been successful in applying analytics techniques to raw sensor data from participant homes to inform clinical partners about the long-term behaviour and behaviour changes in the routine daily in-home lifestyle and activities of participants.  Insights gained from visualising activities at an ADL level rather than at the level of raw sensor data, is more insightful and ultimately beneficial for the participant and the clinician.


\begin{dci}
The Authors declare that there is no conflict of interest.
\end{dci}

\begin{funding}
This work was supported by the Disruptive Technologies Innovation Fund administered by Enterprise Ireland, project grant number DT-2018-0258 and by Science Foundation Ireland under grant number SFI/12/RC/2289\_P2, co-funded by the European Regional Development Fund.
\end{funding}

\noindent 
{\bf Ethical approval:} 

\noindent 
The research ethics committee of Dublin City University approved this study (REC number: DCUREC202221) on 25/1/2022.

\noindent 
{\bf Guarantor:} AFS

\noindent 
{\bf Author Contributions:}

\noindent 
AFS and CMT researched the literature and AFS, CMT, PH, HL and CM conceived the study. CMT, PH and CM obtained ethical approval and performed subject recruitment and HL and HRV managed the data capture and analysis including visualisations.
HRV implemented the ARM coding. 
AFS and CMT wrote the first draft of the manuscript. All authors reviewed and edited the manuscript and approved the final version of the manuscript.

\noindent 
{\bf Acknowledgements}  Not applicable.

\begin{sm}
The raw sensor data from 23 homes used in this work is publicly available on the Figshare repository at \cite{Timon2022}.
\end{sm}



\end{document}